\documentclass[10pt, conference,a4paper]{IEEEtran}
\IEEEoverridecommandlockouts
\usepackage{cite}
\usepackage{amsmath,amssymb,amsfonts}
\usepackage{algorithmic}
\usepackage{graphicx}
\usepackage{textcomp}
\usepackage{xcolor}
\usepackage[utf8]{inputenc}

\usepackage{url}
\usepackage{subcaption}
\usepackage{graphicx}
\usepackage{scrextend}
\usepackage[colorlinks=true, allcolors=blue]{hyperref}
\usepackage{tikz}
\usepackage{tkz-tab}
\usepackage{multirow}
\usepackage{listings}
\usepackage{latexsym}
\usepackage{amssymb}
\usepackage{amsmath}
\usepackage{subcaption}
\usepackage{mathtools}
\usepackage[]{moresize}
\usepackage{soul}
\usepackage{enumitem}   
\usepackage{subfiles}

\usepackage{tabularx,booktabs}
\newcolumntype{C}{>{\centering\arraybackslash}X} 
\newcolumntype{b}{X}
\newcolumntype{s}{>{\hsize=.5\hsize}X}
\newcolumntype{v}{>{\hsize=.3\hsize}X}

\def\BibTeX{{\rm B\kern-.05em{\sc i\kern-.025em b}\kern-.08em
		T\kern-.1667em\lower.7ex\hbox{E}\kern-.125emX}}
		
\usepackage[
top    = 1.9cm,
bottom = 2.90cm,
left   = 1.4cm,
right  = 1.4cm]{geometry}

\begin{document}
	
\title{Intent-Based Orchestration for Application Relocation in a 5G Cloud-native Platform

\thanks{This work has partially been funded by the MARSAL project from EU Horizon 2020 research and innovation programme under grant agreement No 101017171 and from the Spanish MINECO grant TSI-063000-2021-56/TSI-063000-2021-57 (6G-BLUR).}
}

\author{\IEEEauthorblockN{Sergio Barrachina-Muñoz, Jorge Baranda, Miquel Payaró, and Josep Mangues-Bafalluy}
	\IEEEauthorblockA{\textit{Centre Tecnològic de  Telecomunicacions de Catalunya (CTTC/CERCA)}, Barcelona, Spain \\
		\{sergio.barrachina, jorge.baranda, miquel.payaro, josep.mangues\}@cttc.cat}
}

\maketitle


\begin{abstract}

The need of mobile network operators for cost-effectiveness is driving 5G and beyond networks towards highly flexible and agile deployments to adapt to dynamic and resource-constrained scenarios while meeting a myriad of user network stakeholders' requirements. In this setting, we consider that zero-touch orchestration schemes based on cloud-native deployments equipped with end-to-end monitoring capabilities provide the necessary technology mix to be a solution candidate. This demonstration, built on top of an end-to-end cloud-native 5G experimental platform with over-the-air transmissions, shows how dynamic orchestration can relocate container-based end-user applications to fulfil intent-based requirements. Accordingly, we provide an experimental validation to showcase how the platform enables the desired flexible and agile 5G deployments.

\end{abstract}


\section{Introduction} \label{section:introduction}

To embrace 5G and beyond (B5G) visions, the network function virtualization (NFV) paradigm is moving toward the use of containerized or cloud-native network functions (CNFs) to boost scalability, efficiency, and footprint reduction for resource-constrained edge scenarios~\cite{kostopoulos2022experimentation}. Further, there is the need to endow the network with end-to-end monitoring capabilities combining measurements from multiple technological domains (e.g., RAN, transport, or mobile core) to achieve zero-touch close-loop orchestration. Accordingly, this permits the continuous fulfillment of service level agreements (e.g., end-to-end latency) defined as intents to be processed by the orchestrator.

Previous works relying on intent-based approaches~\cite{intent_mdpi_20,intent_velasco_21}, \textit{i)} do not consider a joint solution for full cloud-native systems covering both the 5G core in a multi-PoP environment and end-to-end monitoring (including RAN), or \textit{ii}) only deal with intents covering the mobile entities, i.e., without performing close-loop orchestration actions comprising also user applications.
Instead, this demonstration builds on top of our previous work~\cite{barrachina2022cloud} by prototyping a Kubernetes-empowered intent-based orchestrator that caters to the dynamic lifecycle management of containerized applications to fulfill end-user intents. Remarkably, the orchestrator contemplates end-user and infrastructure provider requirements, paving the way to multi-objective slicing algorithms that must share resources among stakeholders.

\section{System Architecture} \label{section:arch}

\begin{figure*}[ht!]
	\centering
	\includegraphics[width=0.995\textwidth]{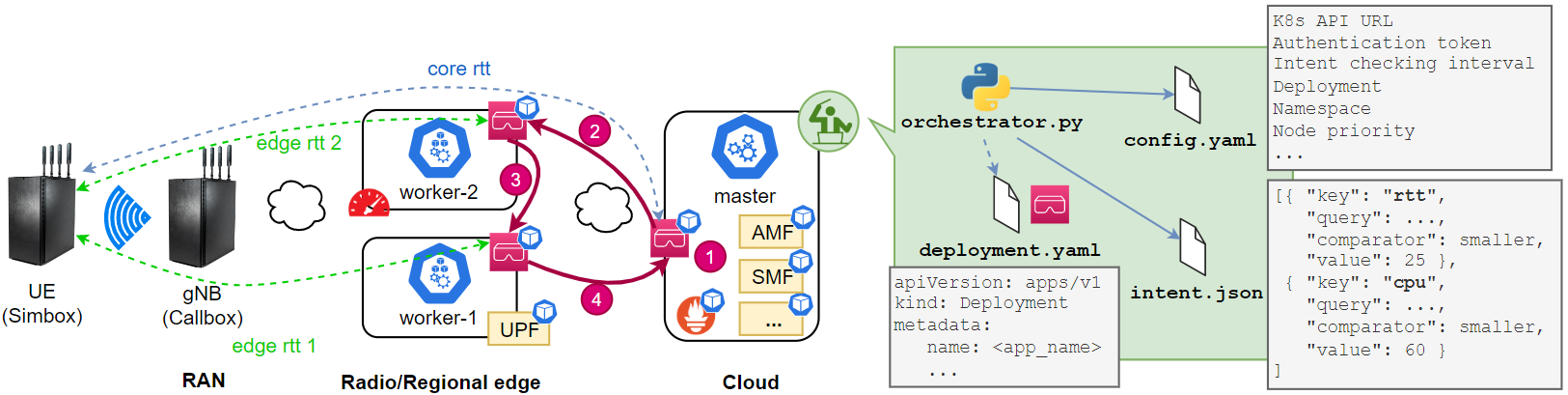}
	\caption{Experimental scenario under demonstration. Experiment steps/events are represented by purple circles.}
	\label{fig:scenario}
\end{figure*}

Fig.~\ref{fig:scenario} illustrates the experimental setup deployed in CTTC's lab premises, featuring the PARADIGMS testbed. The 5G network is composed (from left to right) of the following elements: the Amarisoft UE Simbox Series providing UE emulation capabilities, the Amarisoft Callbox Ultimate acting as a stand-alone 5G gNB, and a Kubernetes cluster consisting of two workers and one untainted master nodes deployed in two points-of-presence (PoPs) representing edge and cloud locations.
The cluster hosts the CNFs constituting the different components of the 5G core (Open5GS v2.4.0), an end-to-end monitoring system, and end-user applications. It is worth mentioning that each 5G Core NF runs as an independent CNF, allowing its distribution between different nodes spanning edge/cloud locations. Indeed, to enable a scenario with a Multi-access Edge Computing (MEC) flavor, we deploy the UPF at the edge rather than in the core. As for the end-to-end monitoring component, we rely on \texttt{kube-prometheus},
 which scrapes metrics of interest exposed by the cluster nodes and gNB and feeds them to the intent-based orchestrator. We refer the reader to~\cite{barrachina2022cloud} for more details.

In this demonstration, the orchestrator is developed as a Python script that relies on the Kubernetes scheduler for materializing its actions. As inputs, the orchestrator receives three files: \textit{i}) orchestrator configuration parameters such as authentication credentials or intent checking interval, \textit{ii}) the UE-APP service requirements defined as a series of conditions within an intent, e.g., the end-to-end latency should be kept below a given threshold, and \textit{iii}) the Kubernetes deployment for the APP. Notice that we can regard each of the inputs to a different stakeholder, e.g., the infrastructure, service, and application provider should take care (at least partially) of the orchestrator configuration, the intent, and Kubernetes deployment, respectively.
Upon request reception, the orchestrator proceeds to the deployment and the continuous life-cycle management of the intent-assigned APP. To do so, it periodically compares the node's statuses to the demanded intent. Should the orchestrator detect the need for relocating the APP from one node to the other upon violation of any intent condition, it triggers such an action through the Kubernetes Python client.

\section{Demonstration} \label{section:demo}

\begin{figure}[ht!]
	\centering
	\includegraphics[width=0.495\textwidth]{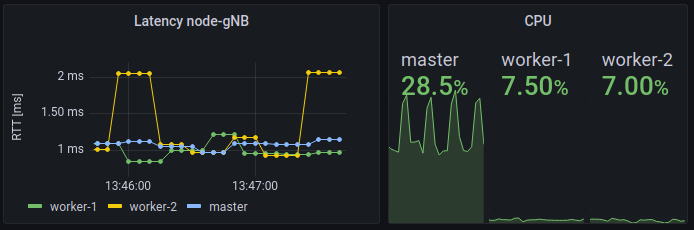}
	\caption{Example of metrics displayed in a Grafana dashboard.}
	\label{fig:grafana}
\end{figure}

This demonstration shows the capabilities of the presented intent-based orchestrator to perform the dynamic lifecycle management of end-user applications. In particular, we emulate the end-user application as an Nginx container, for which, the defined intent imposes certain requirements on the end-to-end latency (from UE to the APP) and on the computing infrastructure performance, meaning that it is needed that the node hosting the application is always below a certain CPU threshold to avoid overloading, and thus grant service reliability. This experiment will be reproduced during the demo, for which we will display a real-time Grafana dashboard (e.g., Fig.~\ref{fig:grafana}), orchestration logs, and the status of the APP container.

\begin{figure}[ht!]
	\centering
	\includegraphics[width=0.495\textwidth]{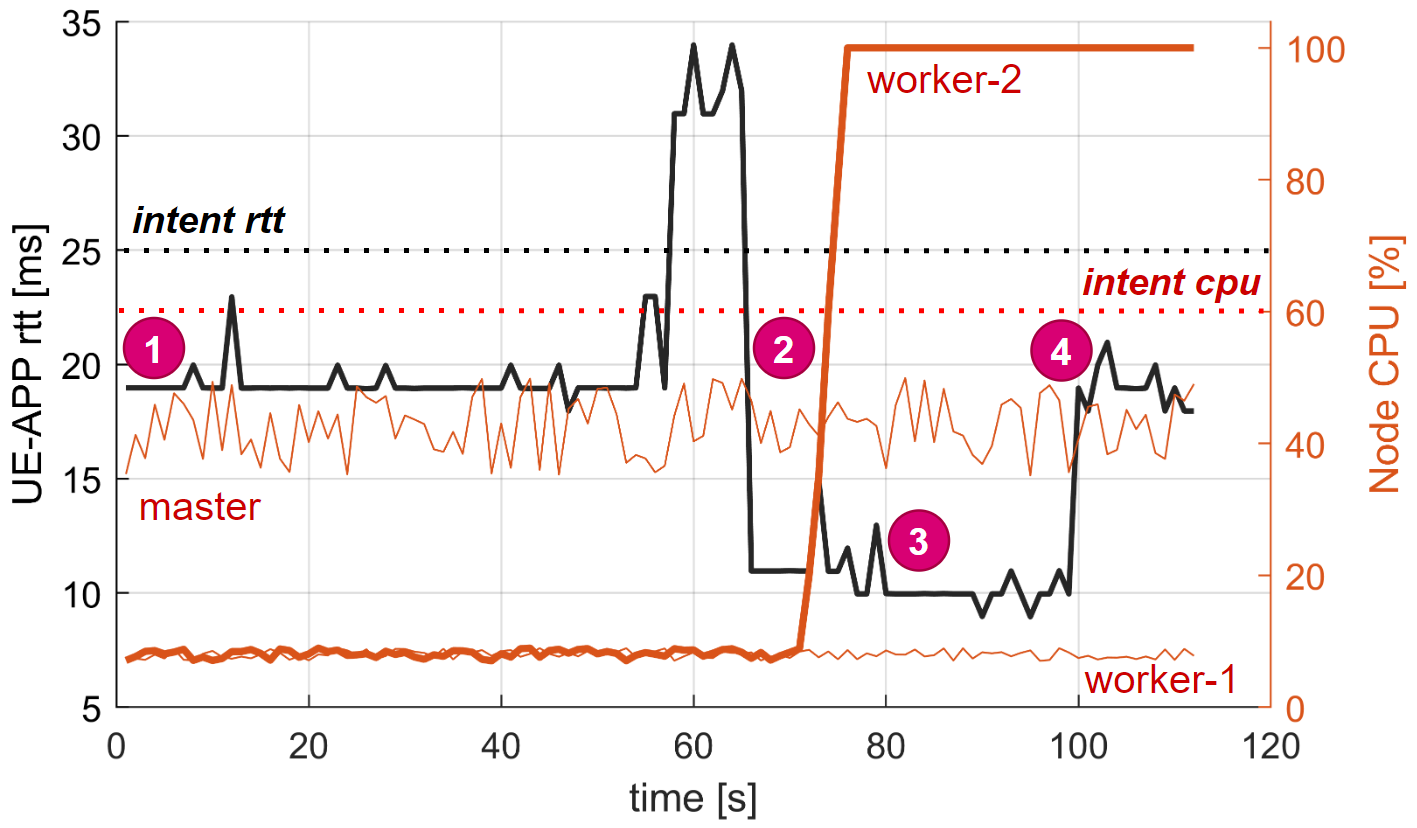}
	\caption{RTT and node CPU usage throughout the demonstration. Experiment events/steps are represented in purple circles.}
	\label{fig:latency}
\end{figure}

The main events of the demonstration, shown in Fig.~\ref{fig:scenario} and Fig.~\ref{fig:latency}, are as follows. In 1), the end-user application is deployed and assigned to an intent requiring a round-trip-time (RTT) from UE to APP below $<$ 25 ms and a CPU usage of the node running the APP below $<$ 60\%. Besides, in the configuration file, we specify region priority for deploying the application. In this example, we assign priority to the master node, assuming that costs are lower in the cloud than in the edge location. Once the APP deployment is ready, the UE can start using the service provided by such an APP. In this case, we run a Python script in the UE side that uses the function \texttt{measure\_latency()} from the \texttt{tcp\_latency} library. 
Notice that in this first stage, the CPU at the master node is higher than in the workers since it runs all the 5G CNFs except the UPF, which runs in worker-1.

In 2), to force an APP relocation, we emulate an addition of 20 ms delay in the network interface of the master node that could be caused by any issue or configuration change in the transport network. We do so with the Ubuntu utility \texttt{tc qdisc}. This also raises the latency from UE to the APP, exceeding the corresponding RTT intent condition. The orchestrator realizes such an issue and checks whether any other node is a valid candidate to deploy the APP. In this case, both worker nodes are and have similar CPU loads, so it randomly selects worker-2 as the new target node and triggers a Kubernetes deployment patch, instantiating the APP deployment in the new node (worker-2) and terminating the APP at the original node (master). Remarkably, this relocation is transparent to the UE since the APP service is kept seamless.

Then, in 3), we raise the CPU usage at edge node worker-2 with the Ubuntu utility \texttt{stress}, so to emulate new workloads generated from other APPs. This way, while the latency condition of the master remains unfulfilled, the CPU usage condition of worker-2 is also violated. Consequently, the orchestrator detects that only edge worker-1 is a valid candidate. Hence, the application is relocated from worker-2 to worker-1. Finally, in 4) the orchestrator detects that the node with priority (master) fulfills the intent conditions and triggers a new relocation to such node accordingly.


\section{Conclusions} \label{section:conclusions}

In this work, we have implemented, tested, and validated an intent-based orchestrator that dynamically relocates APPs to fulfill the requirements of an end-user service, which we materialize as an intent.
As future work, we envision the development and integration of predictors for anticipating the violation of intent conditions, so APPs are proactively relocated aiming to guarantee complete intent or SLA fulfillment.



\bibliographystyle{IEEEtran}
\bibliography{bib}

\end{document}